\documentclass[runningheads]{llncs}
\usepackage{graphicx}

\usepackage{tikz}
\usepackage{comment}
\usepackage{amsmath,amssymb} 
\usepackage{color}
\usepackage{xcolor}
\usepackage{times}
\usepackage{epsfig}
\usepackage{graphicx}
\usepackage{amsmath}
\usepackage{amssymb}
\usepackage{booktabs}
\usepackage{mathtools}
\usepackage{adjustbox}
\usepackage{array}
\usepackage{multirow}
\usepackage{enumitem}
\usepackage{colortbl}
\usepackage{bm}
\usepackage{lipsum}
\usepackage{indentfirst}
\usepackage{mathrsfs}
\usepackage[T1]{fontenc}
\usepackage{bm}
\usepackage{bbding}
\usepackage[section]{placeins}
\usepackage{float}

\usepackage[accsupp]{axessibility}  

\usepackage[pagebackref,breaklinks,colorlinks]{hyperref}

\begin{document}
\pagestyle{headings}
\mainmatter
\def\ECCVSubNumber{58}  

\title{Sliding Window Recurrent Network for Efficient Video Super-Resolution} 

\titlerunning{Sliding Window Recurrent Network for Efficient Video Super-Resolution}
%
\author{Wenyi Lian\inst{1} \and
Wenjing Lian\inst{2}}
\authorrunning{Wenyi Lian and Wenjing Lian}
%
\institute{Uppsala University, Sweden \\ \email{wenyi.lian.7322@student.uu.se} \and
Northeastern University, China \\
\url{https://github.com/shermanlian/swrn}}

\maketitle

\begin{abstract}
Video super-resolution (VSR) is the task of restoring high-resolution frames from a sequence of low-resolution inputs. Different from single image super-resolution, VSR can utilize inter-frames' temporal information to reconstruct results with more details. Recently, with the rapid development of convolution neural networks (CNN), the VSR task has drawn increasing attention and many CNN-based methods have achieved remarkable results. However, only a few VSR approaches can be applied to real-world mobile devices due to the computational resources and runtime limitations. In this paper, we propose a \textit{Sliding Window based Recurrent Network} (SWRN) which can be real-time inference while still achieving superior performance. Specifically, we notice that video frames should have both spatial and temporal relations that can help to recover details, and the key point is how to extract and aggregate these information together. Address it, we input three neighboring frames and utilize a hidden state to recurrently store and update the important temporal information. Our experiment on REDS dataset shows that the proposed method can be well adapted to mobile devices and produce visually pleasant results. 

\end{abstract}

\section{Introduction}

Over the past decade, we have seen the great success of Deep Learning (DL) and Convolution Neural Networks (CNNs) in computer vision~\cite{goodfellow2016deep,lecun2015deep}. By incorporating recent advanced CNN architectures in video super-resolution (VSR), many methods have achieved remarkable performances compared with traditional approaches. Thanks to the deep learning, these improvements encourage the community to explore more solutions for VSR such as sliding window-based methods~\cite{jo2018deep,tian2020tdan,wang2019edvr} and recurrent-based networks~\cite{sajjadi2018frame,haris2019recurrent,isobe2020video,isobe2020revisiting}. These CNN-based methods usually require high computational resources and inference times, and the performance gains mainly come from their huge parameters and complexity. 

On the other hand, with the growing popularity of built-in smartphone cameras, applying VSR networks to real-world mobile devices becomes vitally important and has drawn great attention~\cite{li2022ntire}. However, running CNN models on a smartphone is difficult due to the limited memories and the real-time inference requirement~\cite{ignatov2021real,ignatov2021real2}.  Compared with single image super-resolution (SISR), VSR usually needs to recover a sequence of inter-related frames, and the widely used techniques (e.g., frame alignment and fusion) are too complicated and computationally expensive thus they can hardly be used for smartphones directly.
Moreover, the number of frames can linearly affect the inference times which further aggravates the deployment of CNN-based VSR methods. Some researchers treat VSR as an extension of SISR so that they could use efficient SR architectures without considering the temporal information~\cite{liu2021evsrnet}. Such a solution can achieve real-time inference but shows inferior performance.

To promote the development of real-time VSR, \textit{Mobile AI \& AIM 2022 Real-Time Video Super-Resolution Challenge}~\cite{ignatov2022vsr} is held to evaluate VSR networks on mobile GPUs which have strong resource-constraints. All participants were asked to design efficient models and train them on the REDS~\cite{nah2019ntire} dataset with $4\times$ video upscaling. And participants should design models considering the balance between high restoration accuracies (PSNR) and low resource consumptions (latency). To evaluate the model efficiency, all solutions are asked to convert to `\textit{tflite}' models and tested on the Android \textit{AI Benchmark application}~\cite{ignatov2018ai} which uses the Tensorflow TFLite library as a backend for running all quantized deep learning models. 

\begin{figure}[t]
    \centering
    \includegraphics[width=1.\linewidth]{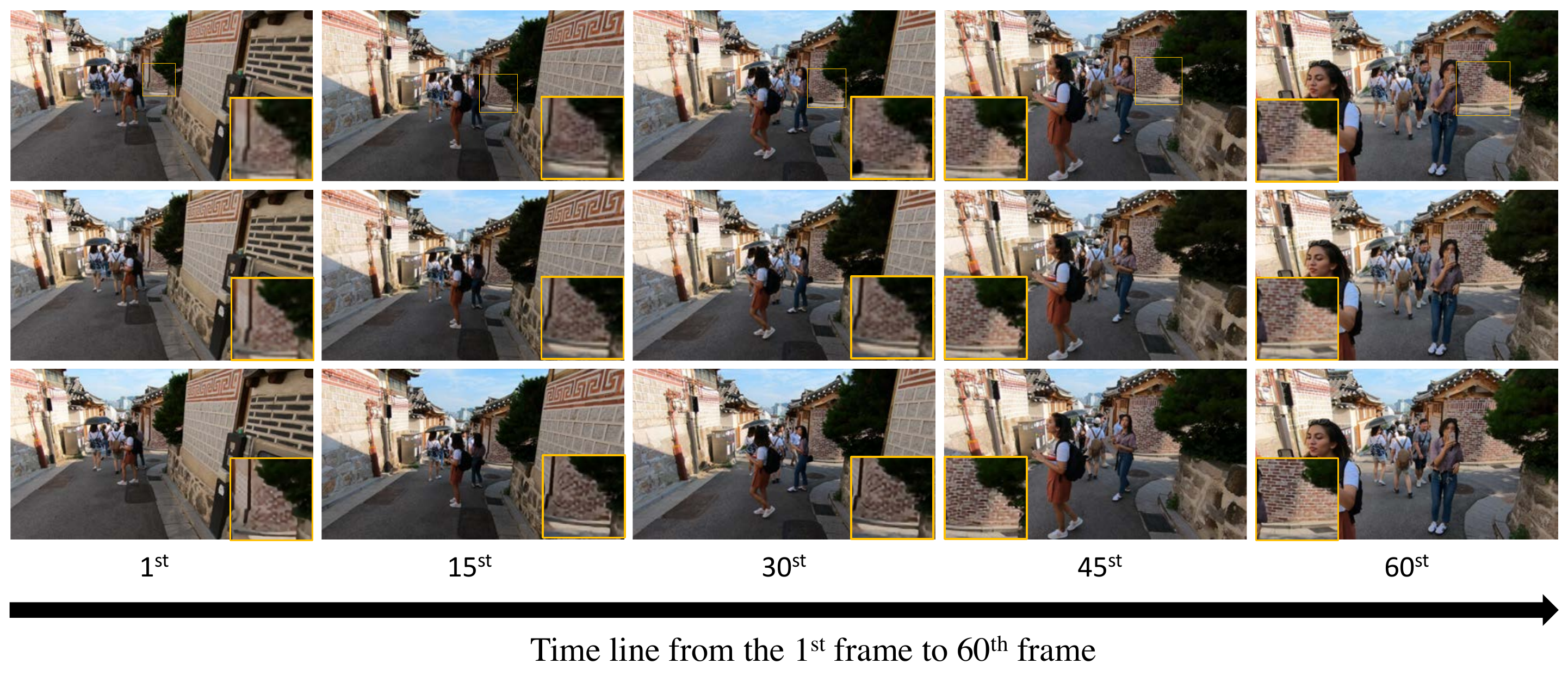}
    \caption{Video sequence from the 1st frame to 60th frame. The top row is the low-resolution frames and the middle row shows the bicubicly upsampled frames. Our method is illustrated in the bottom row. Note that the same part of different frames could have different sharpness and details, which enables us aggregate neighboring frames to recover an HR image with rich details.}
    \label{fig:teaser}
\end{figure}

In this paper, we aim to design a lightweight VSR model by comprehensively investigating the effectiveness of different CNN architectures for smartphones. To reduce the parameters and memories, our network is designed to contain only 3$\times$3 convolution layers and ReLU activation functions. Besides, we find that the same part of different frames could have different sharpness and details,  as illustrated in Figure~\ref{fig:teaser}. To improve the accuracy, we further propose the sliding window recurrent network (SWRN) which makes use of the information from neighboring frames to reconstruct the HR frame. And an additional bidirectional hidden state is used to recurrently collect temporal spatial relations over all frames. By doing so, images produced by our model could have rich details far beyond other single image super-resolution methods. The main contribution of our work can be summarised as follows:

\begin{itemize}
    \item[-] We propose a lightweight video super-resolution network, known as \textit{SWRN}, which can be easily deployed on mobile devices and perform VSR in real-time.
    \item[-] To make the SWRN more efficient while preserving high performance, we propose the sliding-window strategy to utilize neighboring frames' information to reconstruct rich details.
    \item[-] A bidirectional hidden state is incorporated to recurrently store and update temporal information, which could be very useful to aggregate long-range dependencies to improve the VSR performance.
\end{itemize}

\section{Related Works}

\subsection{Single Image Super-Resolution}

Single Image Super-Resolution (SISR) is the task of trying to reconstruct a high-resolution image from its degraded low-resolution low-quality counterpart. In the past few years, numerous works based on deep learning and deep convolution neural networks have achieved tremendous performance gains over traditional super-resolution approaches \cite{dong2014learning,dong2015image,hui2018fast,kim2016accurate,ledig2017photo,luo2022deep,wang2018esrgan,zhang2018residual,lim2017enhanced}. 
SRCNN~\cite{dong2014learning} is the first work that uses CNN in super-resolution. Later, most subsequent works focus on optimizing the network architectures \cite{dong2015image,lai2017deep,zhang2018residual,lim2017enhanced} and loss functions \cite{johnson2016perceptual,ledig2017photo,wang2018esrgan,lugmayr2020srflow}.

\subsection{Video Super-Resolution}
Starting from the pioneer network SRCNN~\cite{dong2015image}, deep convolution neural network based methods have brought significant achievement in both image and video super-resolution tasks~\cite{kim2016accurate,lim2017enhanced,zhang2018image,ahn2018fast,wang2019edvr,liang2021swinir,bhat2021deep,luo2021ebsr,luo2022bsrt,lian2021kernel,chan2021basicvsr}. Particularly, in video super-resolution, where the most important parts are frame alignment, many advanced techniques have been developed to improve the accuracy. For example, VESPCN~\cite{caballero2017real} and TOFlow~\cite{xue2019video} propose to use optical flow to align frames. TDAN~\cite{tian2020tdan} and EDVR~\cite{wang2019edvr} point out that estimating an accurate flow for occlusion and large motion frames is difficult and they choose to align frames using deformable convolution~\cite{dai2017deformable,zhu2019deformable}. Especially, EDVR enjoys the merits of implicit alignment and its PCD module uses a pyramid and cascading architecture to handle occlusion and large motions. Another line of VSR methods also incorporates recurrent networks in video process~\cite{sajjadi2018frame,isobe2020video,haris2019recurrent,isobe2020revisiting}, they usually use the hidden state to record the important temporal information.
For the frame reconstruction part, residual blocks~\cite{he2016deep} and attention mechanism~\cite{wang2018non} are widely used to improve the performance. And recent transformer-based methods~\cite{liang2021swinir,liang2022vrt} also have shown attractive in image/video restoration.

\subsection{Efficient Super-Resolution}

Although most CNN-based approaches can obtain remarkable results in video super-resolution, their performance gains are often up to the network capacities~\cite{ignatov2021real,li2022ntire}. It means that they usually require huge memories and computational resources, which makes it difficult to apply these state-of-the-art models on real-world smartphones that have constrained resources and inference times. Thus the network that is deployed on mobile devices should take care of the particularities of mobile NPUs and DSPs~\cite{ignatov2021real,ignatov2021real2}. To deal with it, \textit{AI Benchmark} application~\cite{ignatov2018ai,ignatov2019ai} is designed to allow researchers to run neural networks on the mobile AI acceleration hardware. Based on the \textit{AI Benchmark} application, \textit{Mobile AI 2021}~\cite{ignatov2021real} and \textit{Mobile AI \& AIM 2022 Real-Time Video Super-Resolution Challenge} is held to promote the development of real-time mobile VSR networks. In this paper, we'd like to follow the setting of these challenges and design an efficient yet mobile-compatible network for smartphones.

\section{Method}

\begin{figure}[t]
    \centering
    \includegraphics[width=1.\linewidth]{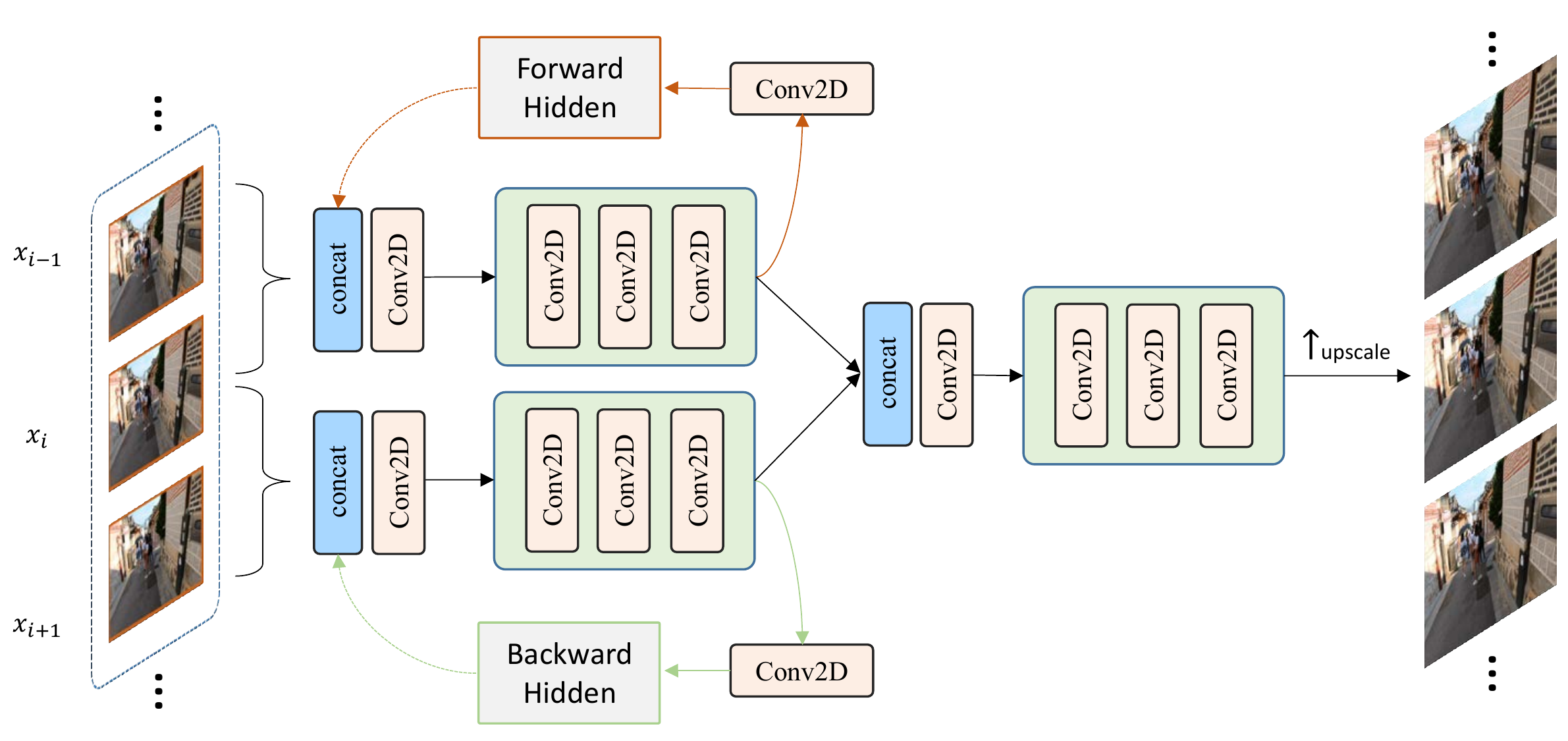}
    \caption{Overview architecture of the proposed Sliding Window Recurrent Network. The forward and backward hidden states are recurrently updated and concatenated with neighboring two frames to provide extra information to reconstruct HR images.}
    \label{fig:network}
\end{figure}

In this section, we introduce basic concepts of video super-resolution and provide a detailed description of the main techniques and strategies of the proposed SWRN for efficient video super-resolution. 

\subsection{Sliding-Window Recurrent Network}
\label{sec:swrn}
The core idea of our method is to grasp and aggregate complementary information from neighbouring frames to improve the performance of video reconstruction. As illustrated in Figure \ref{fig:network}, given the low-resolution frame sequence $\{x_i\}_{i=1}^N, x_i \in \mathbb{R}^{h\times w \times c}$, at each time step $t$, our network accepts three basic inputs (including previous frame $x_{i-1}$, current frame $x_i$ and future video frame $x_{i+1}$) and output a high-quality frame $y_i$, which seems like a sliding window multi-frame super-resolution algorithm~\cite{wang2019edvr,luo2022bsrt}, given by
\begin{equation}
\label{eq:sr}
    y_i = f(x_{i-1}, x_i, x_{i+1}; \theta),
\end{equation}
where $f$ is the VSR network and $\theta$ represents its learnable parameters.
Inspired by~\cite{isobe2020revisiting}, we take the advantage of recurrent hidden states to preserve previous and future information. Specifically, the initial hidden states (forward and backward) are set to 0 and will be updated when the window slides to the next frames. 
Then we can reformulate Equation (\ref{eq:sr}) as follows:
\begin{equation}
\label{eq:hidden}
    y_i, (h_{i+1}^+, h_{i+1}^-) = f(x_{i-1}, x_i, x_{i+1}, h_i^+, h_i^-; \theta),
\end{equation}
where $h_i^+$ and $h_i^-$ are $i$-th frame's forward hidden state and backward hidden state, respectively.
Note here previous frame $x_{t-1}$, current frame $x_t$ and forward hidden state are concatenated as a forward group, then future frame $x_{t+1}$, current frame $x_t$ and backward hidden state compose the backward group. Deep features for each group are separately extracted and concatenated to aggregate multi-frame information to reconstruct the HR frame as:
\begin{align}
\centering
    fea_i^+ &= f_1(concat(x_{i-1}, x_{i})),\\
    fea_i^- &= f_2(concat(x_{i+1}, x_{i})),\\
    output_i &= f_3(concat(fea_i^+, fea_i^-)),
\label{eq:net3}
\end{align}
where $f_1$ and $f_2$ are the NN extractors that learn to obtain forward and backward features, respectively. Then $f_3$ is the aggregation function that merges all information to get final upscaled frames.
Meanwhile, the extracted features of forward and backward groups will update the corresponding forward and backward hidden states by using two simple convolution layers. Then these hidden states can be used for the next frames.

\subsection{Architecture \& Loss}
In training, we use the robust $L_1$ Charbonnier loss~\cite{lai2017deep} to achieve high-quality video reconstruction, which can be formulated as follows:
\begin{equation}
    {\cal L}_{cb}=\sum ^{N}_{i=1}\sqrt{( f(x_{i-1}, x_i, x_{i+1}, h_i^+, h_i^-) - y_{i})^{2}+\epsilon^2},
\end{equation}
where $N$ is the number of training samples, and $\epsilon$ is fixed to $1\times 10^{-6}$. 

There are totally of 14 convolution layers as shown in Figure~\ref{fig:network}. To save the inference time on the mobile device, all layers of our network only consist of a single $3\times3$ convolution layer with the ReLU activation function. The number of channels for all convolution layers is set to 16. And as described in Section~\ref{sec:swrn}, three additional concatenation layers are used to combine information of frames and hidden states.  Moreover, we take the bilinear upsampled current frame as a low frequency residual connection to improve the restoration accuracy.

\begin{table}[t]
\centering
\begin{tabular}{ccccc}
\toprule
Method  & \quad \!  \#Params \quad \! & \quad \! CPU \quad \! &  GPU delegate \quad     &  \quad \! PSNR \quad \!    \\ \midrule
FSRCNN~\cite{kim2016accurate} & 25K & 420ms & 86ms & 26.75  \\
Mobile RRN~\cite{isobe2020revisiting} & 37K & 803ms & 73ms & 27.52  \\
SWRN(ours) & 43K & 883ms & 79ms & 27.92  \\
\bottomrule
\end{tabular}
\vspace{0.3em}
\caption{The table shows the model runtime and PSNR on the Huawei Mate 10 Pro Smartphone.}
\label{tab:time}
\end{table}

\begin{table}[t]
\centering
\begin{tabular}{lcccc}
\toprule
Method  & \quad \!  \#Params \quad \! & \quad \! CPU \quad \! &  GPU delegate \quad     &  \quad \! PSNR \quad \!    \\ \midrule
Baseline & 24K & 489ms & 51ms & 27.05  \\
+ sliding window & 34K & 711ms & 61ms & 27.48  \\
+ hidden states & 43K & 883ms & 79ms & 27.92  \\
\bottomrule
\end{tabular}
\vspace{0.3em}
\caption{Ablation of the sliding-window and hidden states on the Huawei Mate 10 Pro.}
\label{tab:ablation}
\end{table}

\section{Experiment}

\subsection{Dataset}
Our model is trained on the high-quality (720p) REDS~\cite{nah2019ntire} dataset, which is proposed in the NTIRE 2019 Competition~\cite{nah2019ntire} and is widely used in recent VSR researches. In the Mobile AI \& AIM 2022 Real-Time Video Super-Resolution Challenge~\cite{ignatov2022vsr}, the REDS dataset contains 240 video clips for training, 30 video clips for validation and 30 video clips for testing (each clip has 100 consecutive frames). All low-resolution videos are produced by bicubic downsampling with a scale factor of 4.

\subsection{Implementation Details}
For training, the batch size is 16 and all training LR images are randomly cropped to $64\times64$ patches. The total training iterations are set to 250000, and the number of video frames in the training phase is set to 10 and changes to 100 in the testing phase. We use Adam~\cite{kingma2014adam} optimizer with an initial learning rate of $1\times10^{-3}$, and decrease it by half every 50000 iterations. Our network is implemented using TensorFlow2.6 and Keras framework with a single Titan Xp GPU.

\subsection{Experimental Result}

To evaluate the proposed method on Video Super-Resolution, we compare SWRN with FSRCNN~\cite{kim2016accurate} and mobile RRN~\cite{isobe2020revisiting}. The former is a pioneer lightweight CNN-based network and consists of 7 convolution layers (two 1$\times$1 layers for feature shrinking and expanding) and a transpose convolution layer for upscaling. The latter is a lite version of Revisiting Temporal Modeling (RRN) which is a recurrent network for
video super-resolution to run on mobile. Both our SWRN and FSRCNN are quantized with INT8 mode. We use the Peak Signal-to-Noise Ratio (PSNR) as the evaluation metric, and we will also consider the number of parameters and runtime on different mobile accelerators.

\begin{figure}[t]
    \centering
    \includegraphics[width=1.\linewidth]{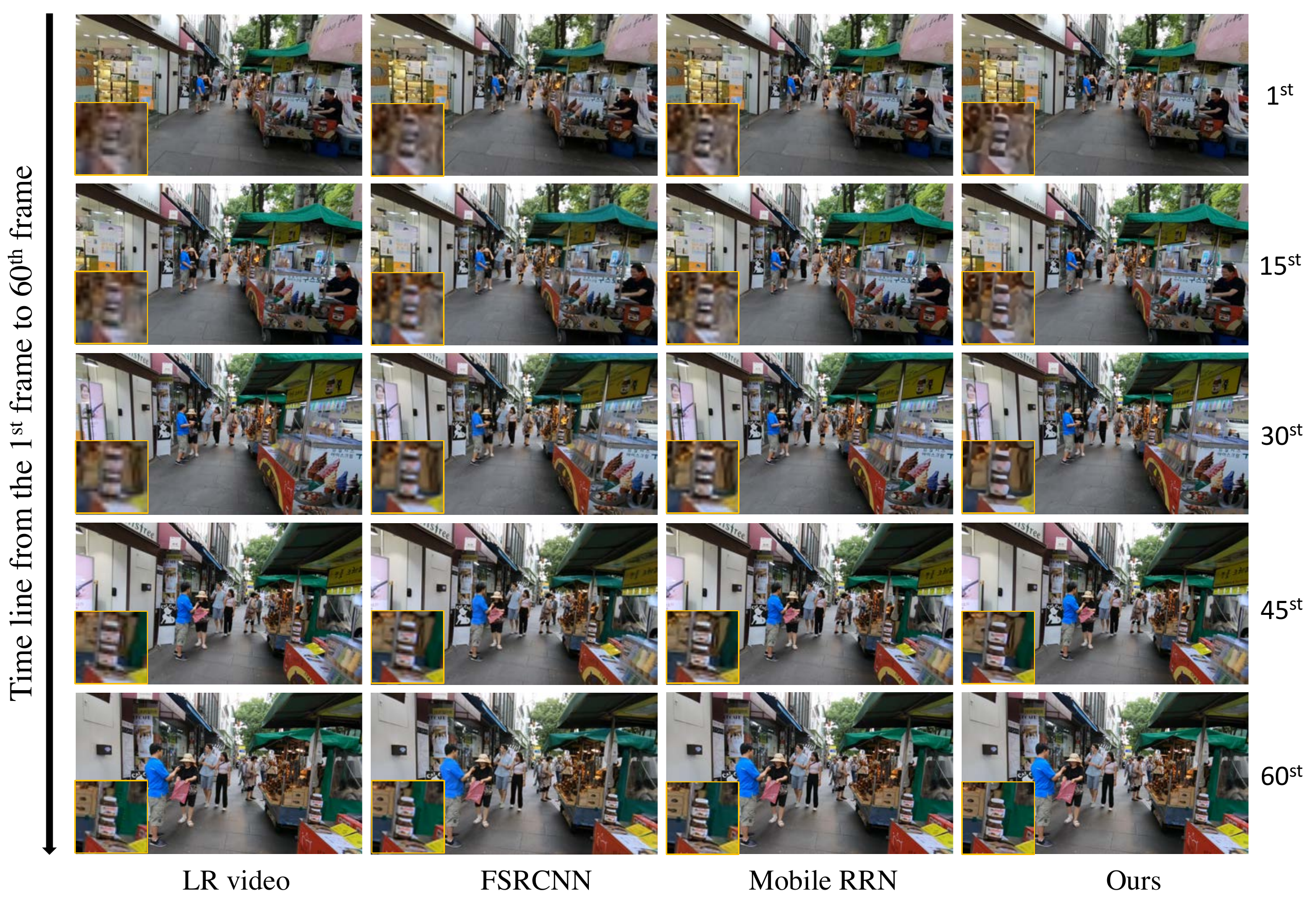}
    \caption{Visual comparison of the video sequence \textit{002} from the 1st frame to 60th frame. }
    \label{fig:compare1}
\end{figure}

\begin{figure}[t]
    \centering
    \includegraphics[width=1.\linewidth]{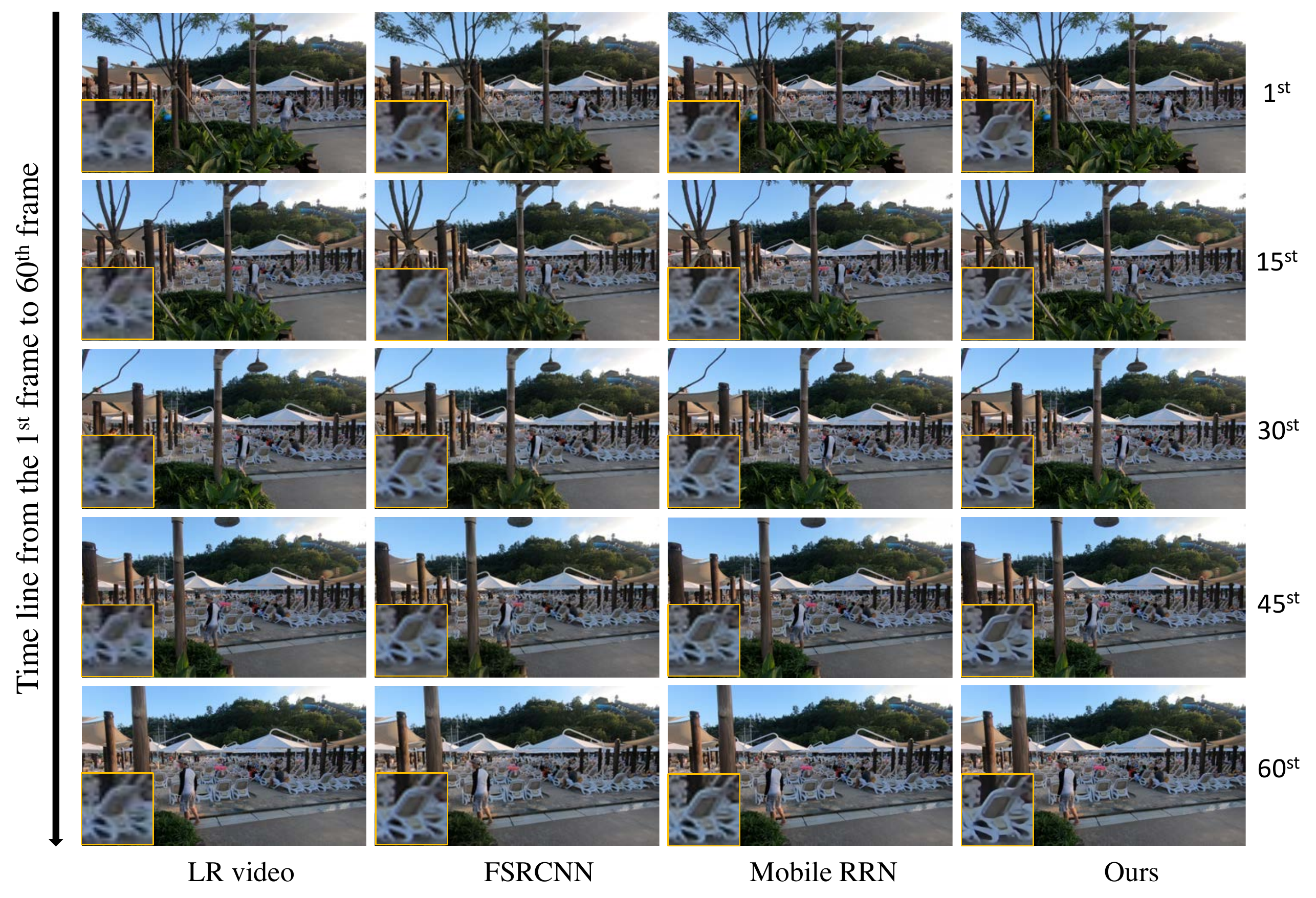}
    \caption{Visual comparison of the video sequence \textit{008} from the 1st frame to 60th frame. }
    \label{fig:compare2}
\end{figure}

The inference times and quantitative results are reported in Table~\ref{tab:time}. One can see that although the proposed SWRN has more parameters and performs slower on the CPU device, SWRN is more compatible with the mobile TFLite GPU delegate accelerator and has lower latency than FSRCNN. In addition, our method achieves a higher performance in terms of PSNR, which surpass FSRCNN by 1.2dB and surpass mobile RNN by 0.4dB. The qualitative results are illustrated in Figure~\ref{fig:compare1} and Figure~\ref{fig:compare2}. One can seen that video sequences produced by our method are much sharper and clear, and are visually pleasant. The result demonstrates our method satisfies the realistic runtime requirement while preserving a high PSNR performance.

\begin{table}[t]
\centering
\begin{tabular}{lcccc}
\toprule
Method  & \quad \!  \#Params \quad \! & \quad \! CPU \quad \! &  GPU delegate \quad     &  \quad \! PSNR \quad \!    \\ \midrule
Channel-8 & 13K & 548ms & 45ms & 26.68  \\
Channel-16 & 43K & 883ms & 79ms & 27.92  \\
Channel-32 & 156K & 2209ms & 232ms & 28.24  \\
\bottomrule
\end{tabular}
\vspace{0.3em}
\caption{Analysis of the impact of number of channels on the Huawei Mate 10 Pro.}
\label{tab:channels}
\end{table}

\subsection{Ablation Study}

In this section, We conduct an ablation study to analyze the impact of the main components of the proposed SWRN framework: sliding-window strategy and (forward and backward) hidden states. For a fair comparison, we use a plain network as our baseline method that only receives the current frame and output a super-resolved image. Moreover, the baseline model uses 3$\times$3 convolution layers with the ReLU activation function, and the number of channels is the same as SWRN. As shown in Table~\ref{tab:ablation}, although the baseline model has fewer parameters and requires lower runtime, it can only achieve 27.05dB in terms of PSNR. By adding the sliding window strategy and hidden states, our method increases the PSNR by nearly 0.9dB, which further demonstrates the superiority of the proposed method. In addition, we also provide the analysis of the impact of the number of channels in Table~\ref{tab:channels}. As we see, although the setting of 8 channels and 32 channels can achieve faster inference time and PSNR performance respectively, the 16-channel setting is more balance in runtime and fidelity metrics.

\section{Conclusion}

In this paper, we propose a \textit{Sliding Window based Recurrent Network} (SWRN) which can be real-time inference on mobile devices while still achieving a superior performance. The basic strategies incorporated are sliding-window and recurrent hidden states. To improve the inference time on smartphones, all layers in our network only contain 3$\times$3 convolution and ReLU activation function. Our method is evaluated on the REDS dataset with a scale factor of 4. The results shows our method is well compatible with mobile TFLite GPU delegate and can run faster than FSRCNN while preserving a high PSNR performance. 

%
%
\bibliographystyle{splncs04}
\bibliography{main}
\end{document}